# Testing the physical reality of tidal bulges in the world's oceans

Yongfeng YANG[1*] Jiajia YUAN[2] Mingyuan FAN[3,4*]

[1] Water Resources Comprehensive Development Center, Bureau of Water Resources of Shandong Province, Jinan 250013, China

[2] School of Geomatics, Anhui University of Science and Technology, Huainan 232001, China

[3] Water Resources Research Institute of Shandong Province, Jinan 250013, China

[4] Shandong Key Laboratory of Water Network Dispatching and Efficient Utilization, Jinan 250013, China

[*]Corresponding authors: Yongfeng YANG (roufeng_yang@outlook.com), Mingyuan FAN (fantina715@126.com) ;

**Abstract**

Persistent alternation of high and low water in coastal and oceanic regions has attracted human attention for millennia. This movement of water is generally explained through the double water bulge model. Although this model has been widely adopted in the scientific literature on tides since the 18th century, the physical existence of water bulges on the Earth's surface has yet to be verified. Herein, we establish a lunar angle phase–dependent statistical analysis of tide patterns at 362,370 oceanic locations spotted by Jason-3 satellite of AVISO in 2021 to address this issue. We show that during lunar angle phases of 0°–45° and 135°–180°, which spatially correspond to the water bulging regions expected in the double water bulge model, the number of low tides consistently exceeds that of high tides. Conversely, during lunar angle phase of 45°–135°, which spatially correspond to the water-depressing region expected in the model, high tides predominantly outnumber low tides. These findings evidently contradict the physical existence of two water bulges in the world's oceans, suggesting that the scientific community should pay additional attention to alternative explanations for tides, such as gravitational forcing mechanism and oceanic basin oscillation-generated driving mechanism.

## 1. Introduction

The regular alternations of high and low water are among the most fascinating natural phenomena on Earth. In most of the world's oceans, two high and two low tides occur each day, whereas in some regions, only one high and one low tides occur per day. Exploring the physics behind this movement of water has tested human wisdom for thousands of years. Many people such as Aristotle, Galileo, Descartes, and Newton—proposed explanations involving the rocky nature of coastlines, water movement induced by the Earth's rotation around the Sun, stresses generated by ether, and the Moon's gravitational attraction, respectively (Pugh and Woodworth, 2014). Although Newton's explanation that the Moon's gravitational force raises seawater is rather simple, it was subsequently developed by others (e.g., Euler, Bernoulli, Laplace, Lord Kelvin, Jeffreys, and Munk), ultimately forming modern theories (e.g., the equilibrium tide and the dynamic tide) (Cartwright, 1999; Robert, 2008; Pugh and Woodworth, 2014).

A double water bulge model (Fig.1) has been widely adopted to explain tidal patterns in physics, oceanography, and geography textbooks (e.g., Tipler and Mosca, 2008; Halliday and Resnick, 2013; Pugh and Woodworth, 2014; Nagle and Guinness, 2017; Garrison and Ellis, 2017; Tarbuck et al., 2019; Zhang, et al., 2025). In this model, two symmetric water bulges are positioned on opposite sides of the Earth's surface. As the Earth spins on its axis, any given location passes beneath these bulges of water, experiencing two high tides and two low tides each day. Extending this idea, when lunar-induced water bulges combine with those generated by the Sun, they either reinforce each other to form spring tides or cancel each other to form neap tides. Such bulge-based explanations are also introduced on the websites of authoritative institutions such as the National Oceanic and Atmospheric Administration of the UAS (https://oceanservice.noaa.gov/education/tutorial_tides/tides03_gravity.html), National Geographic

(https://education.nationalgeographic.org/resource/cause-effect-tides/), and the National Aeronautics and Space Administration (NASA) (https://science.nasa.gov/resource/tides/). However, despite their widespread use, the physical existence of these water bulges has remained unverified since the time of Colin Maclaurin, who first proposed the concept in the 18th century. In fact, most tide researchers do not believe that the water bulges truly form, given the influences of landmasses, basin topography, the Coriolis force, bottom friction, and other factors. Nevertheless, the view has remained largely theoretical or oral, and no observational evidence has been presented.

Thanks to advancements in satellite altimetry over the past 50 years, our understanding of ocean tides has greatly improved (Stammer et al., 2014). Orbital satellites measure sea surface height using laser ranging or radar, and the collected observations contain periodic tidal signals. Through a series of complex procedures (including data corrections and harmonic analysis), tidal data can be extracted. These data can be utilized with a method of double extrapolation to construct ocean tide models, which in turn can reproduce tidal data across global oceans (e.g., Fu et al., 2001; Fok, 2012). In contrast to satellite altimetry, which offers broad coverage of the world's oceans, tide gauges are limited to coastal regions and islands. The tidal records obtained from both altimetry and tide gauges provide a valuable opportunity to address the question of whether water bulges exist. This thus becomes the first objective of this study, with the results expected to provide new insights into the physics of tides. The close correlation between tides and the Moon (Sun) is well established. For instance, large tides often coincide with full and new moons, while small tides occur during the first and last quarters. However, a spatial linkage between the number of high or low tides and the lunar (solar) angle remains unclear. Therefore, investigating this linkage constitutes the second objective of this study.

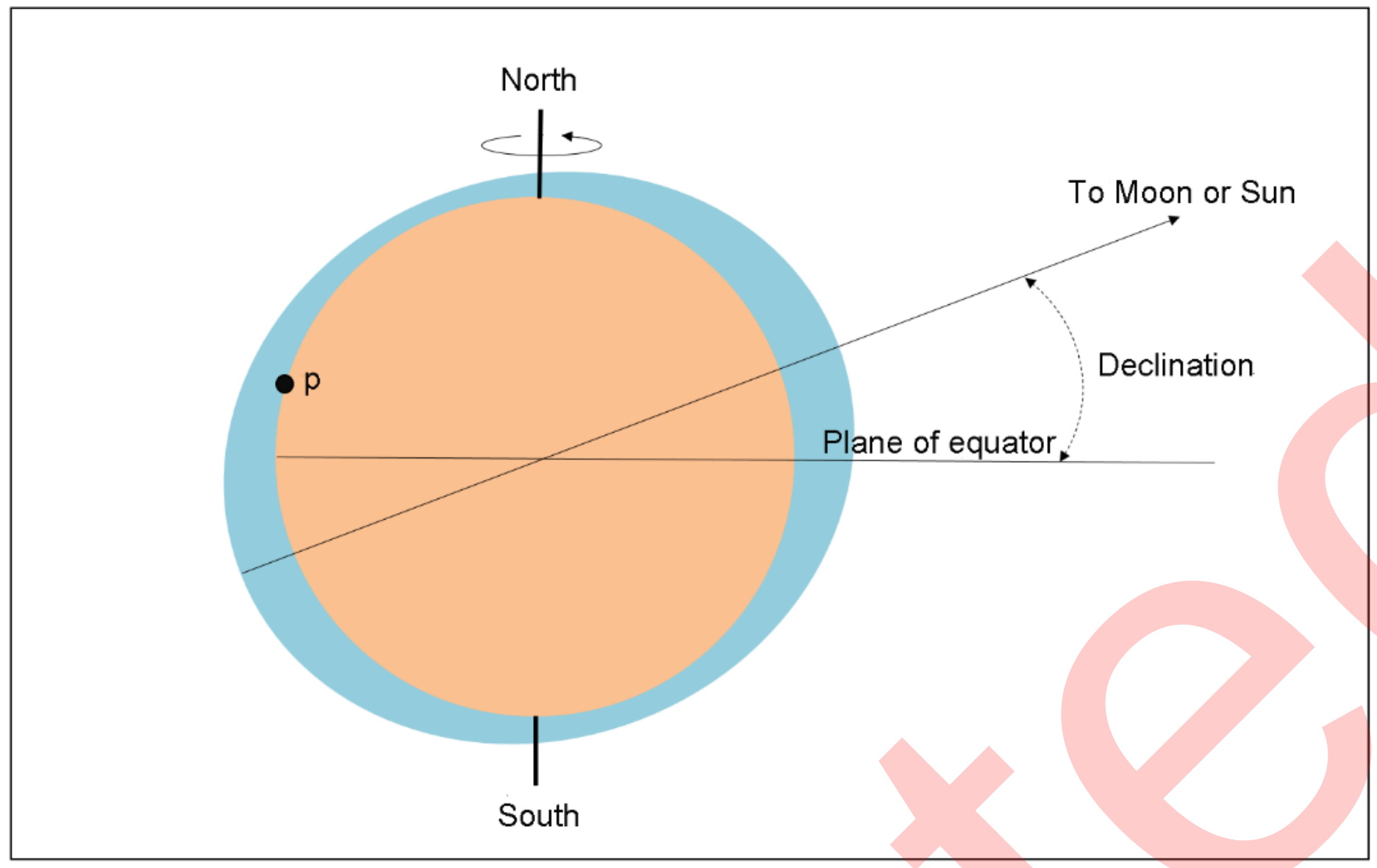

Fig.1 Double water bulge model used to explain tidal patterns. The map is reproduced from Pugh and Woodworth (2014).

## 2. Method and Result

2.1 Method

The spherical symmetry of the double water bulge model allows us to develop an approximate representation of it. We assume that the solid Earth is entirely covered by water and that the two water bulges are aligned with the Moon. As shown in Fig.2(A), segments *a–bf* and *d–ce* represent the two bulging water regions, while *b–c–e–f* represents the depressed water region. Geometrically, if the lunar angle of a location (denoted as *S*) — that is, the angle between the location and the Moon relative to Earth's center — is close to 0° or 180°, it indicates that the location lies within one of the two bulging water regions and thus tends to experience a high tide. Conversely, if the lunar angle is close to 90°, it indicates that the location lies within the depressed water region and thus tends to experience a low tide. Here, we define a high tide as a water level greater than zero and a low tide as a water level less than zero. Consequently, the lunar angle of a location and its tidal pattern (high tide or low tide) at a given time can be correlated. The number of high and low tides can be counted through tidal data. The lunar angle of a location (denoted as $\theta$) can be calculated through spherical trigonometry (Smart, 1940): $\cos\theta = \sin\sigma\sin\delta_{m} + \cos\sigma\cos\delta_{m}\cos C_{mm}$, where $\sigma$, $\delta_{m}$, and $C_{mm}$ are the geographic latitude of location $S$, the declination of the Moon, and the hour angle of the location with respect to the Moon, respectively. The declination of the Moon can be obtained from ephemeris data (i.e., the JPL Horizons system of NASA). Because ocean water is continuous, if we conceptually arrange continents onto an Earth entirely covered by water (as shown in Fig.2(A)), our analysis must necessarily be limited to ocean regions. As more than 70% of the Earth's surface is covered by ocean, the continents can, in a fluid-dynamical sense, be treated as submerged obstacles. Although coastal topography profoundly influences tidal patterns by bending waves in shallow waters, its effect is minimal in the deep open ocean, where depths generally exceed several kilometers. Consequently, continental landmasses are unlikely to exert a dominant influence on the large-scale spatial pattern of the global ocean.

Satellites now provide extensive coverage of the globe. For satellites such as the Jason series of AVISO, the terrestrial regions that they spot are between 66°N and 66°S. Owing to a special orbital design of the satellite, it can only spot one location on the sea surface per second, and next second it moves a few kilometers forwards to spot another location. Therefore, the geographical distance between two adjacent locations spotted along the satellite's ground track is about 7.2 km. Herein, we use minutely tidal data at these locations spotted by Jason-3 of AVISO, and these data are reproduced by Fes2014b ocean tide model (Carrere et al., 2016). Notably, these data represent predictions from the Fes2014b tidal model, rather than raw satellite measurements. The selected dataset spans from 2021-01-01 00:08 to 2021-12-31 23:59. The ground tracks of the Jason-3 satellite (Fig.2(B)) provide global coverage every 10 days; thus, a duration of 365 days represents ~36 rounds of coverage over the Earth's surface. From this dataset, a total of 362,370 data points (one data point per location) between 66°N and 66°S were finally extracted. This large number of tidal observations is sufficient to detect the spatial structure of ocean water on a global scale.

Tide-gauge stations are extensively distributed across coastal regions and islands. The data obtained from these stations are of high quality and publicly accessible through archives such as the University of Hawaii Sea Level Center (UHSLC) and Permanent Service for Mean Sea Level (PSMSL). For this study, we selected hourly tide-gauge data from 166 stations located between 45°N and 45°S for the period from 1 to 30 August 2014, obtained from the database of PSMSL (Caldwell et al., 2015). The selection was made in early January 2017. At that time, 189 stations within this latitudinal belt were included in the PSMSL database; however, 20 of these stations had incomplete data for this period, and three others were located in the Mediterranean Sea and the Red Sea. Therefore, a total of 166 stations were ultimately selected. The geographic distribution of these stations is shown in Fig.2(B). Since the Moon's declination varies mainly between 18°N and 18°S, the chosen latitudinal belt (45°N–45°S) ensures that the tides observed at these 166 stations can capture potential signals of the bulges and depressions of water. As each tide gauge has its own reference water level, the PSMSL data are not referenced to zero water level. To standardize the data, we averaged the hourly water levels at each station over the month and then subtracted this mean from the original data to obtain the final time series, thereby defining zero as the reference water level. The resulting time series from these 166 stations comprise a total of 119,448 hourly data points, approximately 720 data points per station.

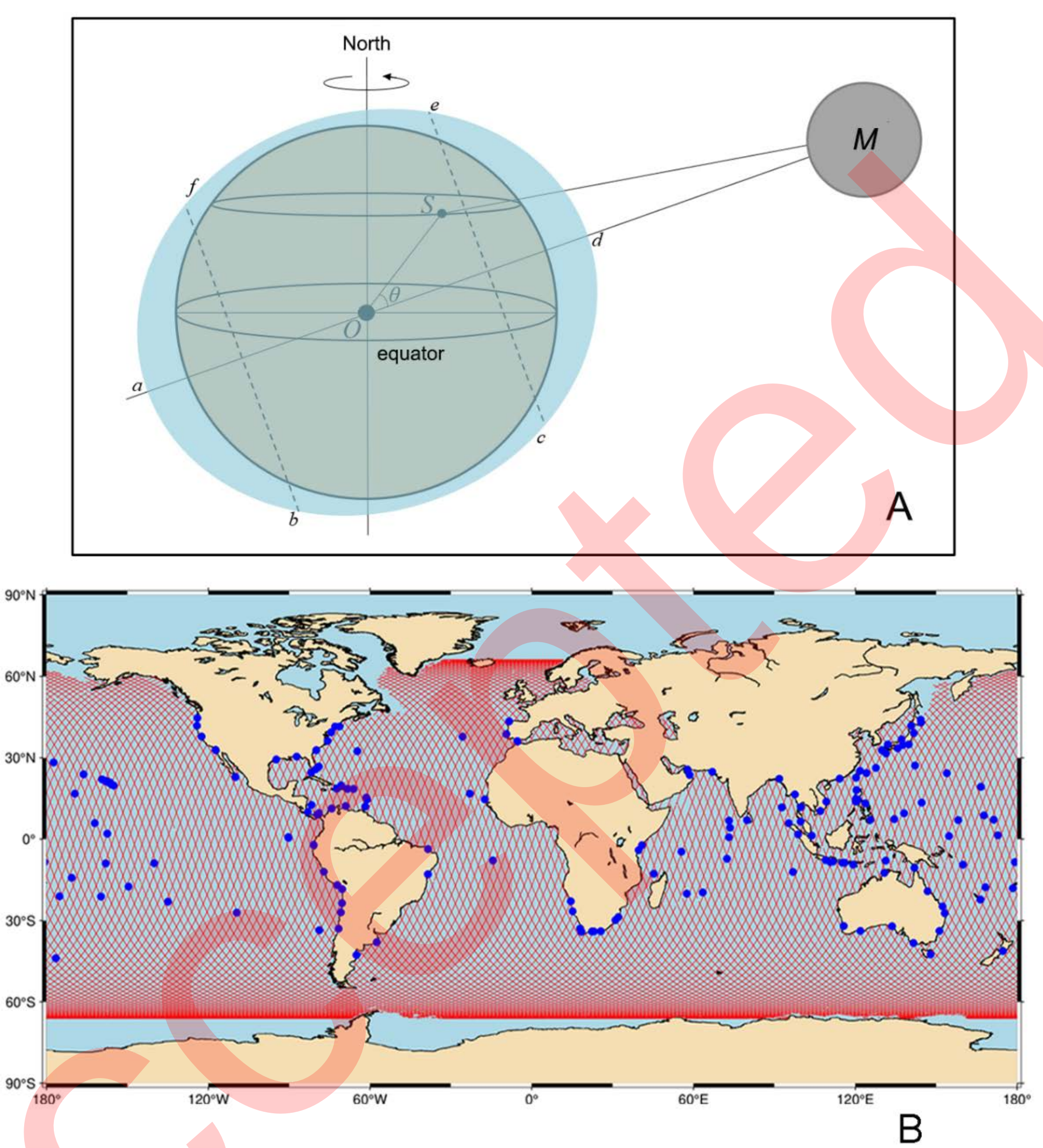

Fig.2 Conceptual model of the double water bulge (A) and the ground tracks of the Jason-3 satellite along with 166 tide-gauge stations (blue dot) (B). Segments *a–bf* and *d–ce* represent two spherical shells corresponding to the bulging regions of water, while *b–c–e–f* is a spherical ring corresponding to the depressed region of water. The ground tracks of the Jason-3 satellite illustrate coverage cycles over 10 days.

## 2.2 Result

We computed the lunar angles of these oceanic locations at the times they were spotted by the satellite and counted the number of high and low tides at these locations. The results are presented in Table 1 and illustrated in Fig.3(A1). We found that during the 0°–60° and 120°–180° lunar angle phases, the number of low tides was consistently greater than that of high tides. By contrast, during the 60°–120° lunar angle phases, the number of high tides consistently exceeded that of low tides. Specifically, there were 87,409 oceanic locations in the 0°–60° lunar angle phase, 186,968 oceanic locations in the 60°–120° lunar angle phase, and 87,993 oceanic locations in the 120°–180° lunar angle phase. Of the 175,402 oceanic locations in the 0°–60° and 120°–180° lunar angle phases, 99,691 (56.84%) experienced low tides, while the remaining 75,711 (43.16%) experienced high tides. Among the 186,968 oceanic locations in the 60°–120° lunar angle phases, 105,418 (56.38%) experienced high tides, with the remaining 81,550 (43.62%) experiencing low tides.

A similar pattern is typically exhibited for the Pacific and Atlantic Oceans (Fig.3(A2 and A3)). In the Pacific Ocean, the number of high tides peaked within the 75°–105° lunar angle phase, while the number of low tides predominated in the 0°–60° and 120°–180° lunar angle phases. In the Atlantic Ocean, the number of high tides peaked at the 75°–90° lunar angle phases. Totally, 168,431 locations in the Pacific Ocean and 113,230 in the Atlantic Ocean were counted during the year 2021. The ratio of the number of locations in the two sets is close to the size ratio of the two oceans.

As a supplement, we investigated the distribution of high and low tides at 166 tide-gauge stations across lunar angles in August 2014, with the results shown in Fig. 3(B). Similarly, low tides predominantly occurred during lunar angle phases of 0°–45° and 135°–180°, whereas high tides were most frequent during the 45°–135° phase. Of the 119,448 data points extracted from these 166 tide-gauge stations, ~56.6% in the former phases corresponded to low tides, compared to 53.5% for high tides in the latter phase.

Table 1 Statistical distribution of high and low tide counts at oceanic locations spotted by the Jason-3 satellite across lunar angles in 2021

| Lunar angle (°) | High tide (number) | Low tide (number) |
|---|---|---|
| 0–15 | 1,849 | 3,272 |
| 15–30 | 6,682 | 9,305 |
| 30–45 | 11,843 | 16,478 |
| 45–60 | 17,425 | 20,555 |

| 60–75 | 23,384 | 21,080 |
|---|---|---|
| 75–90 | 29,435 | 19,937 |
| 90–105 | 29,280 | 19,749 |
| 105–120 | 23,319 | 20,784 |
| 120–135 | 17,232 | 20,596 |
| 135–150 | 12,072 | 16,756 |
| 150–165 | 6,745 | 9,506 |
| 165–180 | 1,863 | 3,223 |
| Total | 181,129 | 181,241 |

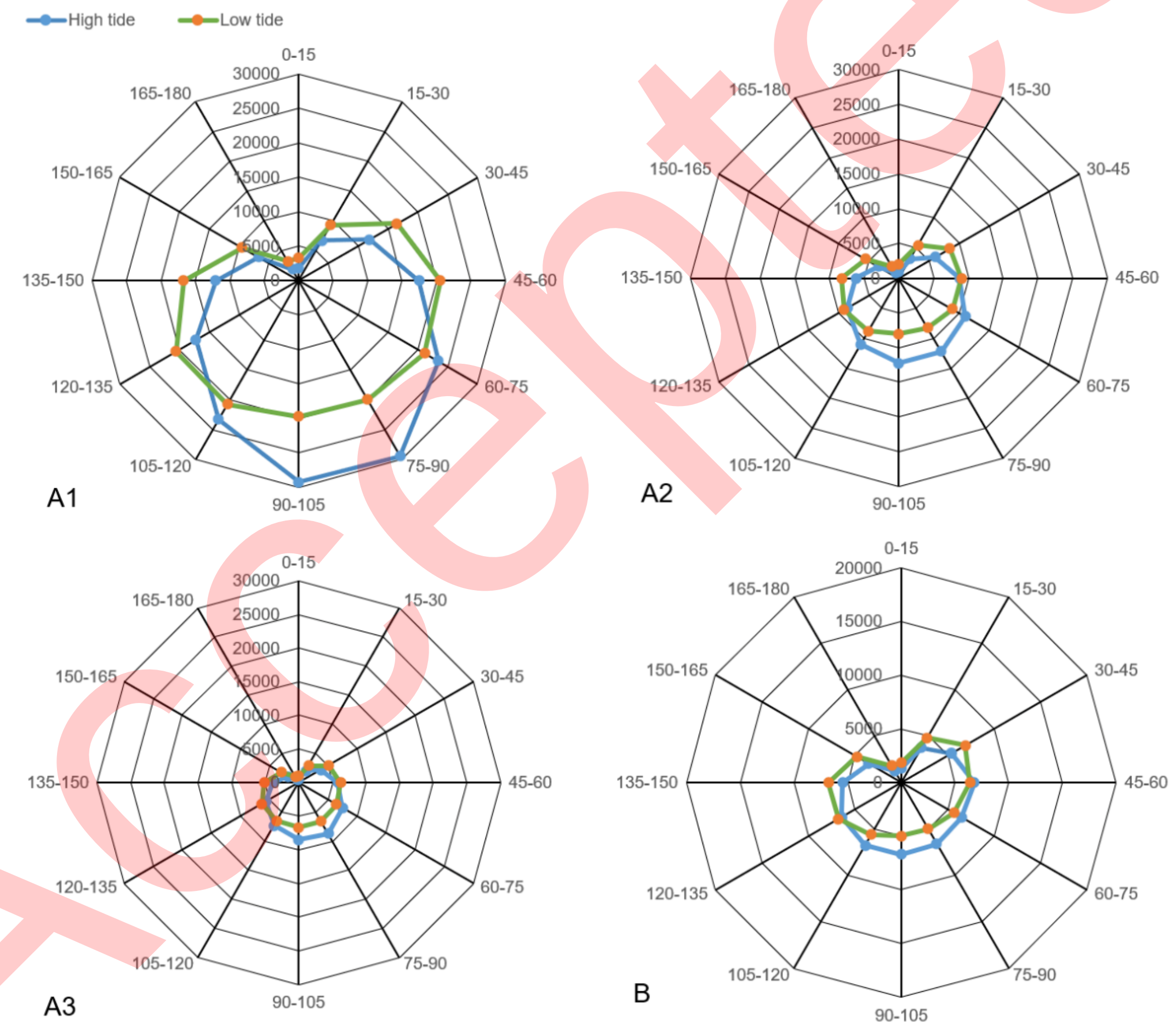

Fig.3 Number of high and low tides versus the lunar angle. Patterns derived from satellite altimetry are shown for the global oceans (A1), the Pacific Ocean (A2), and the Atlantic Ocean (A3). This altimetric dataset includes 362,370 data points (one-minute resolution) from oceanic locations worldwide during 2021. Patterns derived from tide-gauge records are shown in (B), comprising 119,448 hourly data points from 166 tide-gauge stations in August 2014.

To determine a statistical significance of the apparent disparity in the number of high and low tides, a Chi-squared test for independence was conducted throughout the 362,370 observed tide events listed in Table 1. We found a significant effect of lunar angle on the relative frequency of high and low tides, $\chi^2(11) = 4{,}219.6$, $p < 0.001$. This result allows us to conclude that the observed distribution of high and low tides is not owing to chance and is significantly related to the lunar angle.

**3. Discussion**

The patterns reported in Section 2.2 exhibit a feature of stable monthly variations. Low tides predominantly occurred during the 0°–60° and 120°–180° lunar angle phases throughout the year (Fig.4(A and C)). Moreover, the peak activity for low tides in these angle phases occurred from May to June, followed by a significant decline during July–August, and a secondary increase from September to November before decreasing again toward the end of the year. Conversely, high tides predominantly occurred in the 60°–120° lunar angle phase for 11 months of the year (Fig.4(B)), except for November. The peak activity for high tides in this lunar angle phase occurred from January to April, followed by a significant decline during the mid-year (May–July), and a secondary increase from August to October before decreasing again toward the end of the year. This temporal evolution in the distribution of high and low tide number across lunar angles is likely related to the position of the Sun because around late March and late September, the Sun is near the equator, whereas around late June and late December, it reaches its maximum declination.

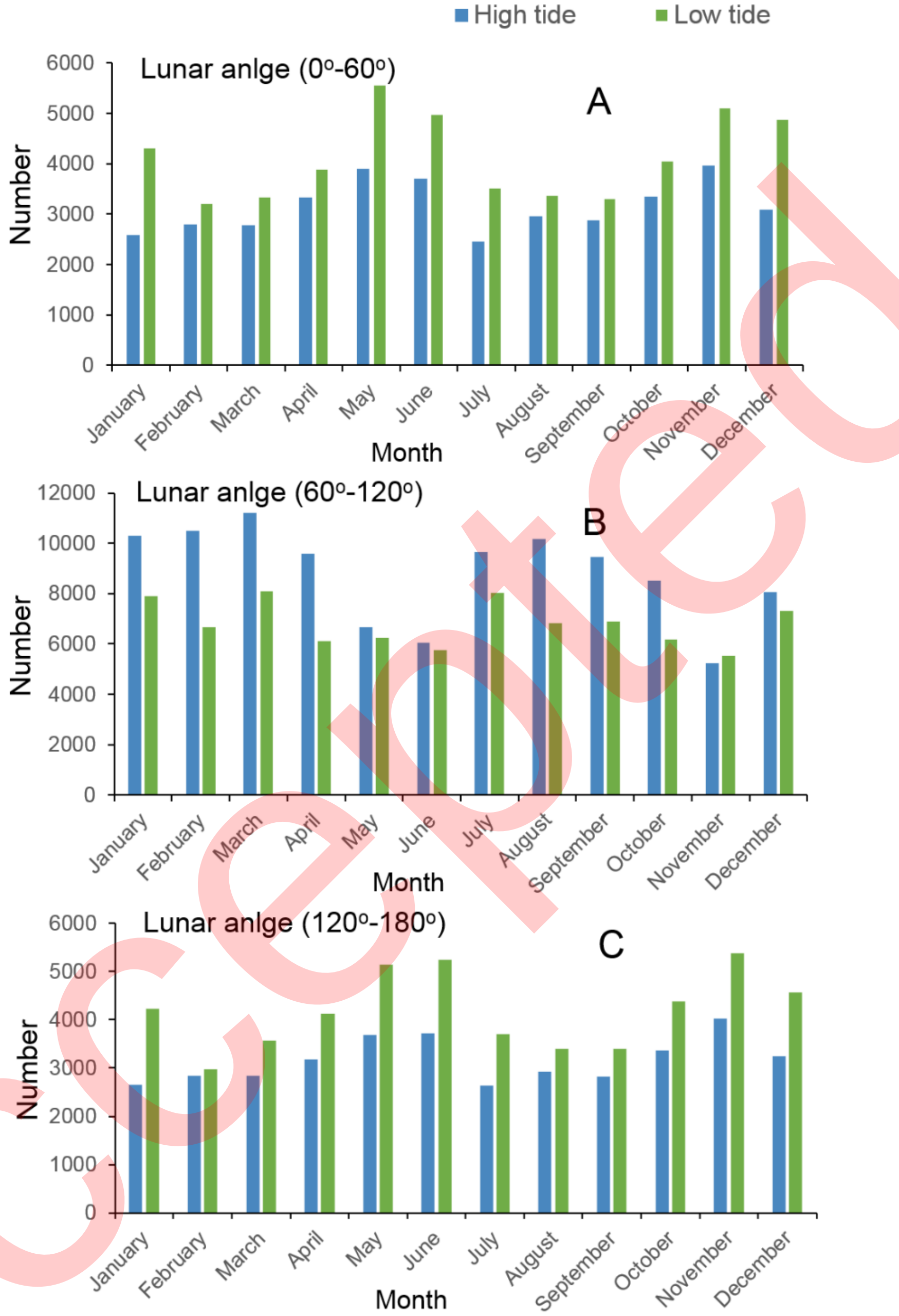

Fig.4 Temporal evolution of number of high and low tides across lunar angles throughout 2021.

In Section 2.2, we presented a correlation between the number of high and low tides and the lunar angle. However, it is widely known that tides are related to both the Moon and the Sun. To gain further insight, we computed the solar angles for the 362,370 oceanic locations spotted by the Jason-3 satellite and counted the corresponding high and low tides. The solar angle was computed using a method similar to that used for the lunar angle, as described in Section 2.1. The results, listed in Table 2 and illustrated in Fig.5, reveal a pattern strikingly similar to that shown in Fig.3(A): low tides occurred predominantly during the 0°–60° and 120°–180° solar angle phases, whereas high tides occurred predominantly during the 60°–120° solar angle phase. This consistent pattern indicates that the presented spatiotemporal evolution of high and low tides is not solely a lunar-driven phenomenon; instead, it is governed by a fundamental mechanism that applies equally to both the Moon and the Sun. Nevertheless, a small but notably disparity exists between the two patterns. Specifically, the contrast between high and low tide number is more pronounced in the lunar angle phases (0°–60° and 120°–180°) than in the corresponding solar angle phases. Similarly, the disparity is greater in the 60°–120° lunar angle phase than in its solar counterpart.

Table 2 Statistical distribution of high and low tide counts at oceanic locations spotted by Jason-3 satellite across solar angles in 2021

| Solar angle (°) | High tide (number) | Low tide (number) |
|---|---|---|
| 0–15 | 2,856 | 2,978 |
| 15–30 | 7,767 | 9,101 |
| 30–45 | 13,037 | 15,867 |
| 45–60 | 16,923 | 18,803 |
| 60–75 | 23,961 | 21,967 |
| 75–90 | 26,364 | 22,339 |
| 90–105 | 26,865 | 22,477 |
| 105–120 | 22,806 | 20,845 |
| 120–135 | 17,124 | 18,798 |
| 135–150 | 13,047 | 16,024 |
| 150–165 | 7,629 | 9,050 |
| 165–180 | 2,750 | 2,992 |
| Total | 181,129 | 181,241 |

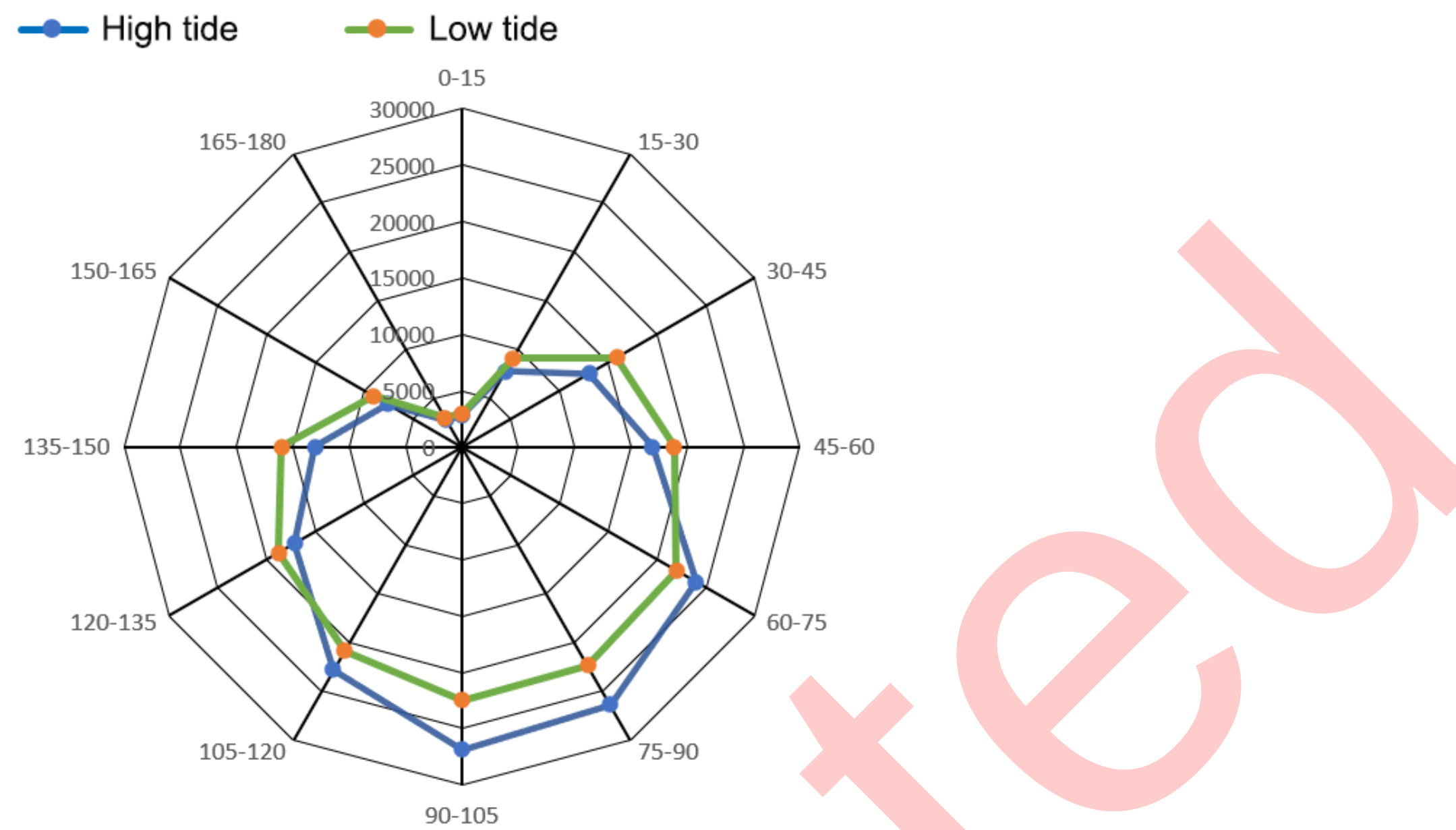

Fig.5 Number of high and low tides versus the solar angle. This dataset includes 362,370 data points (one-minute resolution) from oceanic locations worldwide in 2021.

Our findings on the number of high and low tides across lunar angles directly contradict the physical existence of two water bulges on the Earth's surface. As mentioned earlier, the general public widely accepts the double water bulge model to explain tides, whereas tide researchers prefer the gravitational forcing mechanism. In the forcing mechanism, tides are treated as a manifestation of the response of the complex ocean to tidal forces (Gerkema, 2019). Specifically, as shown in Fig.6(A), the tidal tractive forces (the horizontal components of the tide-generating force) caused by the Moon are assumed to be spherically distributed across the Earth's surface and represented by two symmetric force fields. As the Earth spins, these forces continuously act on the ocean, thereby generating two high and two low tides per day. Extending this concept, when the lunar tractive forces combine with the solar tractive forces, they either reinforce each other to produce spring tides or cancel each other to produce neap tides. However, empirical validation of the effect of tidal forces on ocean water is currently lacking.

Recently, an oceanic basin oscillation–driving mechanism was proposed, where tides are considered a manifestation of oscillations in ocean basins that is intricately linked to the elongated spinning solid Earth due to the Moon (Sun) (Yang, 2025). Specifically, as shown in Fig.6(B), it is assumed that the solid Earth elongates under the gravitational force of the Moon, forming two solid bulges on the Earth's surface. Notably, the magnitude of the solid Earth's elongation had been precisely determined by Yang et al. (2024). As the Earth spins, these solid bulges continuously uplift and depress the ocean basins, inducing water movements that produce two high and two low tides per day. Similarly, when the solid bulges generated due to the Moon combine with those generated due to the Sun, they either enhance each other to make water movement become strongest, producing spring tides, or cancel each other to make water movement become weakest, producing neap tides. This present study provides a robust methodology to validate the effect of the solid Earth elongation on ocean water.

Fluids, such as water, conform to the shape of their container. For example, if one end of a water tank is raised, the water flows toward the other end, becoming shallower at the raised end and deeper at the lower end. Furthermore, if the Earth is assumed to be entirely covered with water, the elongation of the solid Earth would create shallow water above the bulging solid regions and deep water above the compressed solid regions. As shown in Fig.6(C), the spherical shells *a–bf* and *d–ce* represent two shallow regions of water, while the spherical ring *b–c–e–f* represents a deep region of water. Geometrically, if the lunar angle of a location (denoted *S*, for instance) is close to 0° or 180°, it indicates that the location lies within one of the two shallow regions of water and tends to experience a low tide. Conversely, if the lunar angle is close to 90°, it indicates that the location lies within the deep region of water, tending to experience a high tide. The results presented in Section 2.2 are consistent with these expectations: low tides occurred predominantly during the 0°–60° and 120°–180° lunar angle phases, while high tides occurred predominantly during the 60°–120° lunar angle phase. Considering the effect of Sun-induced solid Earth elongation independently, low tides would be expected predominantly along the Earth–Sun line, and high tides predominantly in directions perpendicular to it. Because the solid Earth elongation generated due to the Moon is larger than that generated due to the Sun, the contrast between the number of high and low tides is more pronounced across lunar angle phases than across the corresponding solar angle phases.

Notably, not all oceanic locations within the two shallow regions of water experienced low tides, nor did all oceanic locations within the deep region of water experience high tides. This can be ascribed to the fact that the global ocean is a complex system, in which water movement is determined by multiple factors such as solid Earth deformation, gravitational forcing, wind, atmospheric pressure, and others. Tide lag—conventionally defined as the delay between the Moon's transit and the subsequent high tide—arises from inertia, friction, and the effects of coastal geometry and bathymetry, which create local variations. This delay typically ranges from a few hours to a few days from one location to another. Consequently, the aforementioned patterns in the distribution of high and low tide number across lunar (and solar) angles cannot be fully explained by tide lag alone.

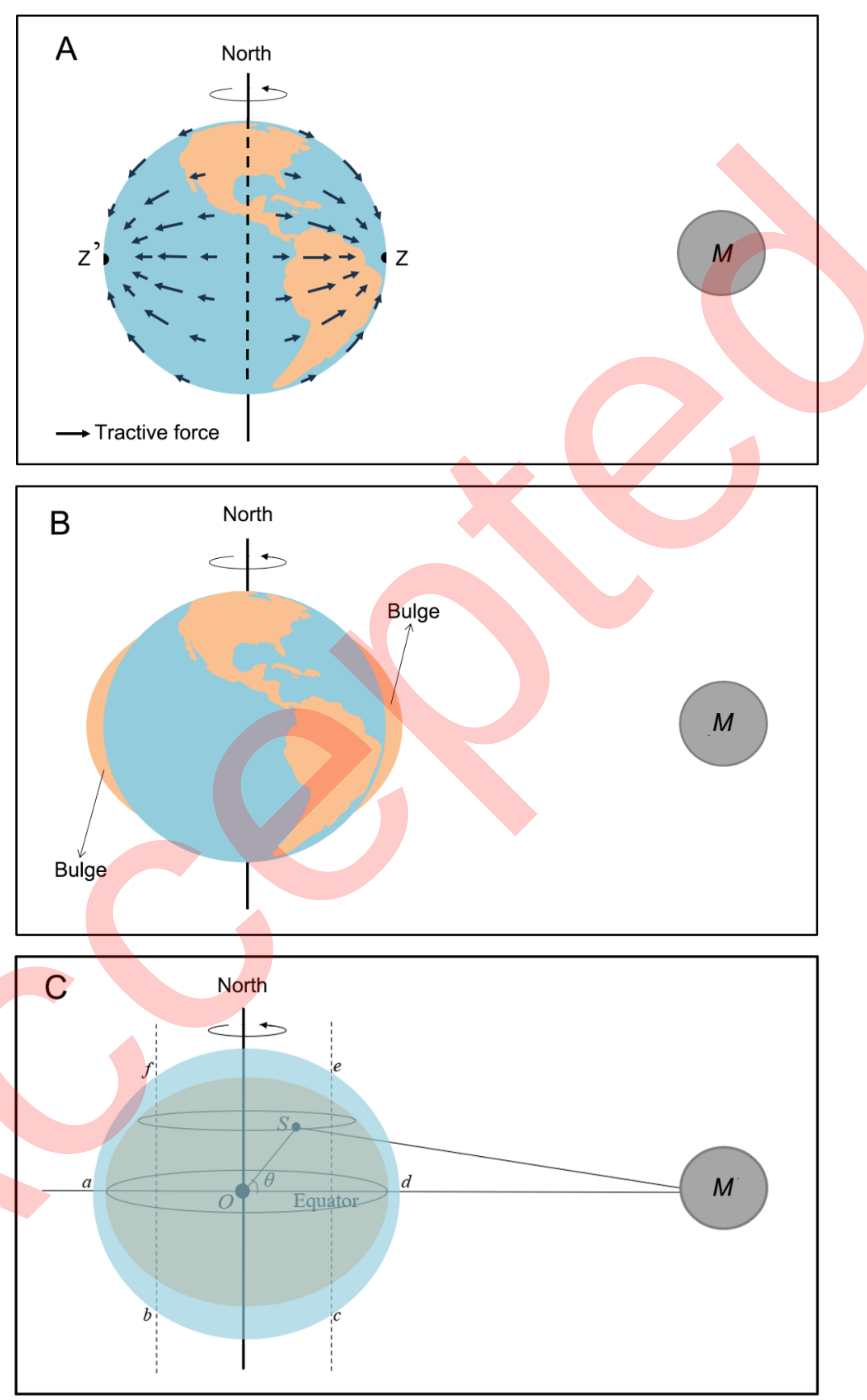

Fig.6 The gravitational forcing mechanism model (A), Oceanic basin oscillating-generating mechanism model (B), and the conceptual model of an elongated solid Earth entirely covered with water (C). Panels (A) and (B) are reproduced from Yang (2025).

**Acknowledgements**

This research was funded by the Key R&D Program Project of Shandong Province (Major Science and Technology Innovation Project) (Grant No. 2023CXGC010905).

**Data Availability**

The altimetry tide data used are available at AVISO (ftp://ftp-access.aviso.altimetry.fr).

The tide gauge data used are available at Permanent Service for Mean Sea Level (PSMSL) (https://psmsl.org/). The ephemeris data used are obtained from NASA JPL horizons system (https://ssd.jpl.nasa.gov/horizons/).